# Cross-Matching Multiple Spatial Observations and Dealing with Missing Data

Jim Gray[1], Alex Szalay[2], Tamás Budavári[2], Robert Lupton[3], Maria Nieto-Santisteban[2], Ani Thakar[2]
1: Microsoft Research, 2: The Johns Hopkins University, 3: Princeton University

**Abstract:**
*Cross-match* spatially clusters and organizes several astronomical point-source measurements from one or more surveys. Ideally, each object would be found in each survey. Unfortunately, the observation conditions and the objects themselves change continually. Even some stationary objects are missing in some observations; sometimes objects have a variable light flux and sometimes the seeing is worse. In most cases we are faced with a substantial number of differences in object detections between surveys and between observations taken at different times within the same survey or instrument. Dealing with such missing observations is a difficult problem. The first step is to classify misses as *ephemeral* – when the object moved or simply disappeared, *masked* – when noise hid or corrupted the object observation, or *edge* – when the object was near the edge of the observational field. This classification and a spatial library to represent and manipulate observational footprints help construct a `Match` table recording both hits and misses. Transitive closure clusters friends-of-friends into object *bundles*. The bundle summary statistics are recorded in a `Bundle` table. This design is an evolution of the Sloan Digital Sky Survey cross-match design that compared overlapping observations taken at different times.

## 1. Terminology: Hits, Misses, Ephemeral, Masked, Edge

Given several observations of the sky, called *runs*, astronomers often want to *cross-match* all the observations of each object from all runs that observed that object. A typical first step is to process the runs to make an object *catalog*. The catalog entries typically take the form:

```
(runID, objectID, position, positionError,
       other attributes…)
```

Two objects are said to *match* if they come from different runs and if their positions differ by less than their *classification distance*.

Picking the classification distance depends on the data and on the intended use of the cross-match. If only stationary objects are to be matched, then the classification distance can be a small multiple of the maximum of the two object's circular rms position errors. The position uncertainty or astrometric precision is often a constant for all objects of an observation, but when comparing data from different instruments or from times with different seeing, the position uncertainties may differ. Various systematic effects can add to uncertainties. A rigorous statistical argument, based on mean density and other parameters can recommend an optimal Bayes classification distance. Given a point in one run, the probability in finding another point at a separation *r* in another run, given perfect accuracy is the sum of a Dirac delta for the object plus the contribution from a spatial correlation function (from clustering) and a random Poisson component. The observational errors, motions, and sizes all create their own errors, which must be convolved with this distribution. These convolutions will broaden the Dirac delta. At the same time there are inevitable false detections and chance overlays. We want a classification distance that minimizes the overall error (i.e. false positives and false negatives.) Ideally one could use a Bayes decision criterion, but the object surface density is not uniform on the sky.

Some studies are interested in moving objects and other studies are working with data collected over an epoch where the earth's observational position affects the object's relative position. In those cases the object's apparent movement may exceed the positional error, and therefore a larger threshold is needed for the match criterion. The technique described here can handle slow-moving objects – where the relative motion during the observational epoch is small compared to the average distance among objects. We return to that issue in Section 5, but for now assume that we only intend to cross-match stationary objects.

For example, SDSS Data Release 5 [6] chose a classification distance of 1.0". The survey has an astrometric precision of 0.1" and an average inter-object distance of 21"; but it chose the high classification distance, 10x the astrometric precision, to include slowly-moving objects in the cross-match. If the SDSS were in the galactic plane, not to mention the galactic center, it would have very crowded fields, and would have a combinatorial explosion using such a large classification distance.

In what follows we assume that the study has selected a classification distance function:
```
ClassificationDistance( positionError1
                      , positionError2).
```

After the coarse spatial match, different astronomers may want to use different morphological and attribute tests to detect spurious matches where a moving object has occluded or changed the attributes of some object or to tease apart adjacent members of a binary system. Having a short list of all candidate match objects allows more sophisticated tests to work much more quickly by limiting their search space.

Given two runs that overlap, if object O1 observed in run1 matches object O2 in run2, we call the pair an O1-run2 *hit*. Indeed more than one object in run2 may match O1, in which case there are several O1-run2 hits. If there are no O1-run2 hits, we call it an O1-run2 *miss*. O1-run2 misses can have three generic causes (see Figure 1):

**Ephemeral**: O1 is at the detection threshold and the seeing was good in run1 but not as good in run2 or O1 may be invisible if run2 is a different kind of instrument (e.g. run1 is optical and run2 is radio or Xray),
or O1 is a variable or transient object which varied below the detection threshold in run2,
or O1 moved more than the classification distance between the two observations.
**Masked**: O1 was fully masked by a meteor trail, cosmic ray, satellite, moving object, passing airplane, or refraction of a bright object in run 2.
**Edge**: O1 was on the edge of the run2 footprint and so not all its pixels were observed.

The three *ephemeral* cases are indistinguishable without a model that captures O1's variability and trajectory. About one third of the primary objects in the SDSS are near the detection threshold, many stars are variable or binary, and supernovae are fairly common in galaxies. In the SDSS about 84% of the match pairs avoid these problems, but 11% of the matches are *ephemeral*, about 0.5% are *masked*, and because the SDSS overlap areas are typically long-narrow strips, about 5% are *edge* objects.

When comparing runs from different instruments the *ephemeral* issues may be even more dramatic – the object may not be visible in the second instrument because it does not radiate in that spectral band, or the two instruments may have very different sensitivity.

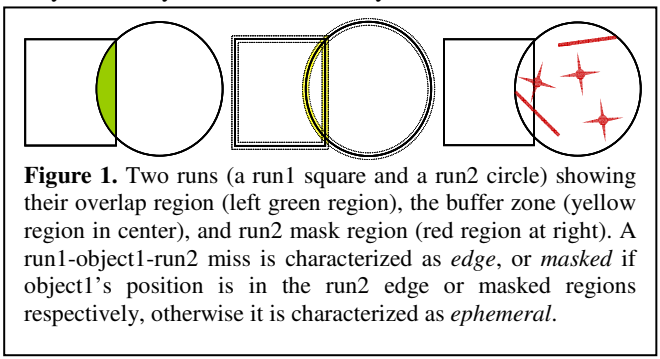

**Figure 1.** Two runs (a run1 square and a run2 circle) showing their overlap region (left green region), the buffer zone (yellow region in center), and run2 mask region (red region at right). A run1-object1-run2 miss is characterized as *edge*, or *masked* if object1's position is in the run2 edge or masked regions respectively, otherwise it is characterized as *ephemeral*.

Summarizing, given two runs, an object in run1 may match (hit) one or more objects in run2, or it may be a miss in run2. Run2 misses may be caused by ephemeral, masking, or edge effects.

Our goal is to compute a table
```
Match( run1, objectID1,
       run2, objectID2,
       hitOrMiss)
```
Where the `hitOrMiss` field takes on one of the values `Hit`, `Ephemeral`, `Masked`, or `Edge`. When the objectID1-run2 pair is a miss, then `objectID2` is zero, and the `hitOrMiss` flag suggests why (`Ephemeral`, `Masked`, or `Edge`)

## 2. Computing Match Hits

Building the `Match` hits from a catalog is easy. In pseudo-SQL:
```
insert Match(run1, objectID1,
             run2, objectID2,
             hitOrMiss)
select Obj1.run, Obj1.objectID,
       Obj2.run, Obj2.objectID,
       'Hit'
from Catalog as Obj1
join Catalog as Obj2
on distance(Obj1.position, Obj2.position)
      < ClassificationDistance(
                    Obj1.positionError,
                    Obj2.positionError)
   and Obj1.run != Obj2.run
```
Indeed, the SDSS catalog pre-computes the spatial join as the `Neighbors` table using the Zones algorithm described in [2]. So the *hit* query is even simpler -- one just looks for neighbors within 1" with run1≠run2 since `Neighbors` stores all object pairs within 30".

## 3. Computing Match Misses

Computing misses is more complex. First we need to know for each object O1 in run1 what other runs overlap O1 to within the run-run2 classification distance. Given such a run2, we need to know if the missing object O1 is either near the run2 footprint edge or is inside a run2 mask. Those two tests characterize the miss as ephemeral, masked, or edge.

Such a test requires precise definitions of the run footprints (spatial extent), and for each run, a list of its masks and their footprints. We adopted the International Virtual Observatory definition for footprints [1] and have implemented a footprint service both inside SQL [2] and on the web [3, 4].

As explained in [2, 4], spherical regions are represented as the union of convex hulls that are each the intersection of a set of half-spaces. A library lets astronomers create regions, do Boolean algebra on them, and do point-in-region tests. This representation dovetails with the HTM library [1] that makes it easy to find all points within a region. Source code for these spatial functions (`buffer, intersect, inside, fRegionGetObjectsFrom-RegionID, ..`) can be found in the SDSS SkyServer implementation available [7].

Given that machinery, it is fairly easy to explain how misses are discovered and characterized. First, using OpenGIS terminology, define `buffer(run1, fuzz)` as a region that expands region run1 by the `fuzz`. Given the run1 region, we need only consider other runs where

```
 intersect( run1
            buffer(run2
                    ,ClassificationDistance)
          ) ≠ Ø
```

If this is an inexpensive test and if there are less than a thousand runs, then one can compute the overlapping run pairs by simply comparing all runs to all others. Otherwise some bounding-box spatial-index is needed to reduce the number of region comparisons. In either case, the computation produces a table

```
Overlap (run1, run2,
         overlapRegionID,
         overlapRegionEdgeID,
         run2MasksID)
```

that records the overlap region of each pair of runs that have a non-null (buffered) overlap. The "edge" region describes the buffer zone (of width:

  `ClassificationDistance(run1.positionError,`
                           `run2.positionError)`,

and `run2MasksID` is the ID of the union of all the mask regions in run2 (see Figure 1.)

Now compute the table of all the misses

```
  Miss(run1, objectID, position1, run2)
```

as follows:

```
insert Miss
select R.run1,C.objectID,C.position, R.run2
from Overlap as R -- overlap region
cross apply fRegionGetObjectsFromRegionId(
                    R.OverlapRegionID)as C
         -- get catalog objects in region
where R.run1 = C.run -- restrict to run1
and not exist (select run1, objectID1, run2
       from Match M -- object not in Match
       where M.run1 = R.run1
         and M.run2 = R.run2
         and M.objectID1 = C.objectID)
```

For each `Overlap` record, this code uses the HTM `fRegionGetObjectsFromRegionId` function to search the catalog for `run1` objects that are in the `run1-run2` overlap region but do not yet have a `run2` entry in the `Match` table.

When this is done, the `Miss` table lists all the O1-Run2 misses. Now we categorize each miss and put that characterization in the `Match` table. First we find the edge cases by:

```
insert Match(run1, objectID1,
             run2, objectID2,
             hitOrMiss)
select Miss.run1, Miss.objectID,
       Miss.run2,          0, 'Edge'
from Miss
join Overlap as O
  on  Miss.run1 = O.run1
  and Miss.run2 = O.run2
where Inside(Miss.position,
             O.OverlapRegionEdgeID)
```

Those `Miss` records can now be discarded by:

```
delete Miss
from Miss
join Match
  on Miss.objectID1 = Match.objectID
 and Miss.run1 = @run1
 and Miss.run2 = @run2
```

Masked misses, use the `Overlap.run2MasksID` which is region ID of the union of the run2 and the HTM code to identify all the Miss objects inside the mask region:

```
insert Match(objectID1,  run1,
             objectID2,  run2,
              hitOrMiss)
select @run1,  Miss.objectID,
       @run2,           0, 'Masked'
from Miss
join Overlap as Masks
  on  Miss.run1 = Masks.run1
  and Miss.run2 = Masks.run2
where Inside(Miss.position,
             Masks.run2MasksID)
```

Those `Miss` records can now be discarded from Miss (using the delete statement above).

The residual misses are neither edge nor masked so they must be ephemeral. They can be added to the `Match` table as

```
insert Match(objectID1,  run1,
             objectID2,  run2, hitOrMiss)
select  run1, objectID,
        run2,         0, 'Ephemeral'
from Miss
```

## 3. Friends-of-Friends – Match Transitive Closure

Matches are not transitive. For example, in Figure 2 object O1 matches O2 and O2 matches O3 but object O1 may not match O3. This might be caused by the object moving, or it might just be an unusually large position error, or they might just be different objects. In any case it is often convenient to group all the friends-of-friends together and treat the whole ensemble as a single group – what we call a *bundle* in the next section.

Computing the friends-of-friends is fairly simple. The match table is grown with the new `hitOrMiss='Friend'` records as follows.

```
-- compute least fixed point of transitive
-- closure.
-- quit when no new rows are added.
until (@@rowcount == 0) {
 insert Match -- add friends of friends
 select distinct M1.run1, M1.objectID1,
                 M2.run2, M2.objectID2,
                 'Friend'
 from Match M1
 join Match M2 -- as transitive closure
   on M1.run2    = M2.run1
  and M1.objectID2 = M2.objectID1
  and ( M1.run1 <> M2.run2--avoid O1=O1
     or M1.objectID1 <> M2.objectID2)
 where not exists (
      select *    -- but skip already
      from Match M -- present edges.
      where M.run1    = M1.run1
        and M.objectID1 = M1.objectID1
        and M.run2    = M2.run2
        and M.objectID2 = M2.objectID2)
 }
```

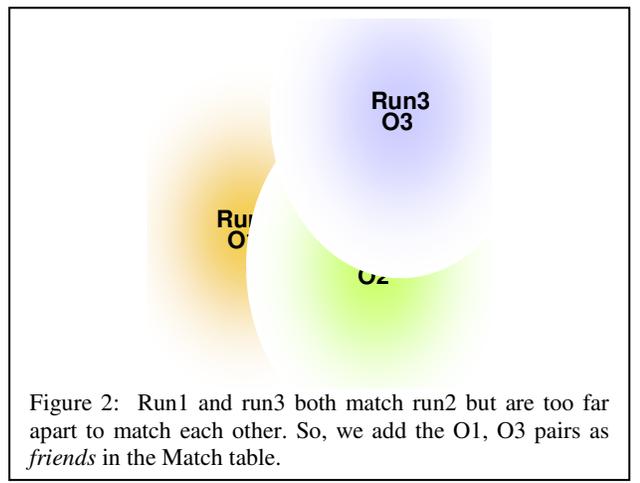

Figure 2: Run1 and run3 both match run2 but are too far apart to match each other. So, we add the O1, O3 pairs as *friends* in the Match table.

## 4. Bundles

Having the `Match` table makes it easy to reason about the observations of the same object and easy to collect statistics (average, variance,…) about the object's position, magnitude, classification, the number and types of misses that the object experienced, and other attributes.

This suggests creating a `Bundle` table that records these statistics.

```
Bundle(bundleID, hits, misses,
       PositionAverage, positionVariance…)
```

Each `Match` record has a `bundleID` field added to it to point to its corresponding `Bundle` record. When bundles overlap it may make sense to merge them into one bundle with one `Bundle` record. As new runs are acquired, new records are added to the catalog and new records are added to the `Match` table (which is easily computed incrementally.) These new records may create new bundles or may add to an existing bundle. One complication is that adding records may cause bundles to merge if the new record causes one bundle to overlap another.

It is easy to compute the aggregate statistics for the bundle table once each match record has an assigned bundle ID. Computing the bundle IDs is a bit tricky so that code is included here.

```
------------------------------------------
-- create a temporary table holding
-- the minimal run, objectID pair
-- in each bundle
create table BundleTemp(
         BundleID int identity primary key,
         run int, objectID int)
-- populate the table with the min elements
insert BundleTemp (run, objectID)
select  run1, objectID1
from Match
where  run1 < run2
   or (run1 = run2
      and objectID1 < objectID2)
group by all run1, objectID1
having count(*) = 0
```

```
-------------------------------------------
-- assign the bundleIDs to each Match table
-- entry
-- that is related to this minimum element
update Match
set BundleID =
 (select R.bundleID
  From (
     select B.bundleID, run2 as run,
            objectID2 as objectID
     from Match M
     join BundleTemp B
       on  M.run1 = B.run
       and M.objectID1 = B.objectID
      union select * from BundleTemp
      ) R
  where Match.run1     = R.run
    and Match.objectID1 = R.objectID)
-- cleanup
drop table BundleTemp
```

## 5. SDSS Experience, Moving Objects, and Multi-Survey Cross Matches.

### 5.1. SDSS Cross-Match Examples

The SDSS catalog is cross-matched with FIRST, RC3, ROSAT, Stetson, and USNO-B as part of the pipeline processing.

About 109M SDSS deblended objects lie in regions observed more than once. These objects cluster into 50M bundles described in the `MatchHead` table of the SDSS DR5. Most bundles are just two observations but about 3M have three observations and 133K have four observations. About 84% of the matches are hits. Of the 16% that are misses, 11% are *ephemeral*, 0.5% are *masked*, and 5% are *edge* because the SDSS overlap areas are typically long-narrow strips.

### 5.2. Moving objects

Most objects are slow-moving so their displacement between observations is small compared to the average inter-object difference. Near-Earth object apparent motions are typically large and so measurements must be within minutes for the techniques described here to detect object pairs. For faint stellar and galactic objects, the apparent motion is typically much smaller and so the observations can be months or years apart and yet the techniques here can correlate the two observations.

The SDSS is observed in five spectral bands – each band's observation occurs about a minute after the previous band. Those five measurements allow cross-matching observations of objects with apparent motions of 0.01 to 10 arcminutes per minute (or a comparable number of degrees per day). The SDSS processing pipeline looks for such objects and records their apparent velocities (in units of degrees per day) in the catalog.

Query 15B of the standard 35 SDSS queries [8] shows how to extend the built-in pipeline cross-match to use the 5-band temporal observations find objects with even greater velocities. That query finds ten additional primary fast-moving objects in Data Release 5.

When considering SDSS observations separated by days or years, only very slow-moving objects can be detected with the cross-match techniques here. For example, in SDSS the average inter-object distance is 21". Given this rather low object surface density (when compared to the Galactic Plane or the Galactic Center), the techniques described here can find slowly moving objects by using a larger classification distance.

But if the object moves more than a few arcseconds per year or if the object density is much higher, then the classification distance technique will hit a combinatorial explosion with too many false-positives.

In general, a naive spatial match does not work for fast-moving objects. Rather one must model the object's motion, and then predict where that object will be in the observational field. Unfortunately, model uncertainties accumulate with time – especially for fast moving near-earth objects. Nonetheless, several surveys (Palomar-QUEST [9], Pan-STARSS [10], LSST [11], and others) are attacking exactly these problems.

## 5.3 Pivoted Cross-Match

The examples discussed so far built match pair tables. Even the SDSS cross-match with FIRST, RC3, ROSAT, Stetson, and USNO-B built pair tables. But, it is sometimes the case that one wants a match table of the form (x, y, z) built from three surveys X, Y, Z where the match elements are the corresponding elements of the survey -- in general the problem involves more than three surveys or observations, but three is enough to demonstrate the issues. For example, the SDSS QSO candidate objects organize the Target, Spectroscopic, and Best cross-match catalog in this way.

Expressed in relational terms this is a *full-outer spatial join* among the N catalogs. The *full-outer part* of that expression means that there may be zero, one, or many items that match for each bundle. If there are no matches in a catalog then that field is filled in with the relational *null* value. At least one column of every row is not null (every bundle has at least one member in one dataset.) In case multiple objects from one catalog qualify, there is usually a "primary" object from that catalog. Often a row containing the primary members is flagged as the primary cross-match of the *N* catalogs.

We call such a cross-match representations a *pivoted cross-match* (as opposed to a pair-table cross-match) because this representation is the *pivot* of the pairs table on the match-head and run number.

Building pivoted cross-matches is surprisingly difficult. A simple strategy is to build the pairs table and bundles as described above and then build the pivoted cross-match as a join from the bundle table. That is what we did for the QsoCatalog table of SDSS DR5 and for a 4-band NDWS pivoted cross-match.

Given the bundle and match tables, the pivoted table can be constructed, using zero rather than null for missing objIDs, as follows:

```
create view  Bundle_Match as
  select distinct bundleID, objID1, run1
  from Match
insert Pivoted (bundleID, x, y, z)
  select B.bundleID,
         X.objID1, Y.objID1, Z.objID1
  from Bundle B
  left outer join Bundle_Match X
    on B.bundleID = X.bundleID
    and run1 = 'X'
  left outer join Bundle_Match Y
    on B.bundleID = Y.bundleID
    and run1 = 'Y'
  left outer join Bundle_Match Z
    on B.bundleID = Z.bundleID
    and run1 = 'Z'
```

## 5. Summary

This approach to classifying and organizing a series of point-source spatial observations addresses the problem faced by astronomers doing a cross-match of multiple runs – either within a survey or between dissimilar surveys. Similar problems arise in other domains. Dealing with missing data is the most difficult problem. The first step is to classifying misses as *ephemeral* – meaning that the object moved or appeared or disappeared or was at the detection threshold, *masked* – meaning that the object was hidden or corrupted by noise in the observation, or *edge* – meaning that the object was near the edge of the observational field. This classification combined with a spatial library to represent and manipulate observational footprints and masks can construct of a Match table recording both hits and misses. The matches can be extended by transitive-closure to friends-of-friends all occupy approximately the same region.

This transitive closure partitions all the observations into disjoint bundles. Information summarizing information about all the observations of an object can then be recorded in a Bundle table. The resulting schema is shown in Figure 3.

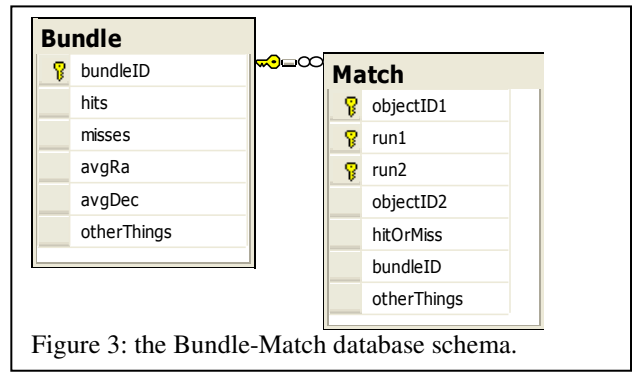

Figure 3: the Bundle-Match database schema.

The design described here evolved from the MatchHead-Match table cross-match implemented for SDSS Data Release 5 [6] and described in [5].